\documentclass[twocolumn,showpacs,aps,prl,superscriptaddress,preprintnumbers,amsmath,amssymb,floatfix]{revtex4}
\usepackage{graphicx}
\usepackage{dcolumn}
\usepackage{bm}

\usepackage{relsize}
\usepackage{xspace}
\def\babar{\mbox{\slshape B\kern-0.1em{\smaller A}\kern-0.1em
    B\kern-0.1em{\smaller A\kern-0.2em R}}}
\def\pep2{PEP-II}
\def\Kp    {\ensuremath{K^+}\xspace}
\def\Km    {\ensuremath{K^-}\xspace}
\def\piz   {\ensuremath{\pi^0}\xspace}
\def\pip   {\ensuremath{\pi^+}\xspace}
\def\pim   {\ensuremath{\pi^-}\xspace}
\def\Dz{D^0}
\def\KS    {\ensuremath{K^0_{\scriptscriptstyle S}}\xspace}
\def\D       {\ensuremath{D}\xspace}
\def\Dp      {\ensuremath{D^+}\xspace}
\def\Dm      {\ensuremath{D^-}\xspace}
\def\Dz      {\ensuremath{D^0}\xspace}
\def\Dzb     {\ensuremath{\Dbar^0}\xspace}
\def\DzDzb   {\ensuremath{\Dz {\kern -0.16em \Dzb}}\xspace}
\def\DpDm    {\ensuremath{\Dp {\kern -0.16em \Dm}}\xspace}
\def\Dstar   {\ensuremath{D^*}\xspace}
\def\Dstarb  {\ensuremath{\Dbar^*}\xspace}
\def\psiprpr  {\ensuremath{\psi(3770)}\xspace}
\def\BR         {{\ensuremath{\cal B}\xspace}}
\def\invfb   {\ensuremath{\mbox{\,fb}^{-1}}\xspace}
\def\Dbar    {\kern 0.2em\overline{\kern -0.2em D}{}\xspace}
\def\Db      {\ensuremath{\Dbar}\xspace}
\def\DDb     {\ensuremath{\D {\kern -0.16em \Db}}\xspace}
\def\DDbX     {\ensuremath{\D {\kern -0.16em \Db}X}\xspace}
\def\DDstarb     {\ensuremath{\D {\kern -0.16em \Dstarb}}\xspace}
\def\DstarDstarb     {\ensuremath{\Dstar {\kern -0.16em \Dstarb}}\xspace}
\def\Y#1S{\ensuremath{\Upsilon{(#1S)}}\xspace}
\def\FourS {\Y4S}
\def\MM{M^2_{rec}}

\newcommand{\ev}{\ensuremath{\mathrm{\,e\kern -0.1em V\!}}\xspace}
\newcommand{\gevc}{\ensuremath{{\mathrm{\,Ge\kern -0.1em V\!/}c}}\xspace}
\newcommand{\mevc}{\ensuremath{{\mathrm{\,Me\kern -0.1em V\!/}c}}\xspace}
\newcommand{\gevcc}{\ensuremath{{\mathrm{\,Ge\kern -0.1em V\!/}c^2}}\xspace}
\newcommand{\mevcc}{\ensuremath{{\mathrm{\,Me\kern -0.1em V\!/}c^2}}\xspace}
\newcommand{\mev}{\ensuremath{\mathrm{\,Me\kern -0.1em V\!}}\xspace}
\newcommand{\gev}{\ensuremath{\mathrm{\,Ge\kern -0.1em V\!}}\xspace}
\newcommand{\BABARPubYear}    {07}
\newcommand{\BABARPubNumber}  {060}
\newcommand{\SLACPubNumber} {12818}

\begin{document}


\preprint{BABAR-PUB-\BABARPubYear/\BABARPubNumber\\}

\preprint{SLAC-PUB-\SLACPubNumber}

\title{\boldmath Study of the Exclusive Initial-State-Radiation Production of the \DDb System.}

\author{B.~Aubert}
\author{M.~Bona}
\author{D.~Boutigny}
\author{Y.~Karyotakis}
\author{J.~P.~Lees}
\author{V.~Poireau}
\author{X.~Prudent}
\author{V.~Tisserand}
\author{A.~Zghiche}
\affiliation{Laboratoire de Physique des Particules, IN2P3/CNRS et
Universit\'e de Savoie, F-74941 Annecy-Le-Vieux, France }
\author{J.~Garra~Tico}
\author{E.~Grauges}
\affiliation{Universitat de Barcelona, Facultat de Fisica,
Departament ECM, E-08028 Barcelona, Spain }
% ...
\collaboration{The \babar\ Collaboration}
\noaffiliation

\date{\today}

\begin{abstract}
A search for charmonium and other new states is performed in a study of 
exclusive initial-state-radiation production of \DDb events from
electron-positron annihilations at a center-of-mass energy of 10.58 \gev.
The data sample corresponds to an integrated luminosity of 384~\invfb
and was recorded by the \babar\ experiment at the \pep2 storage ring.
The \DDb mass spectrum shows clear evidence of the \psiprpr plus 
other structures near  3.9, 4.1, and 4.4 GeV$/c^2$.
No evidence for $Y(4260)\to \DDb$ is observed, leading to 
                          an  upper limit of
$\BR(Y(4260)\to \DDb)/\BR(Y(4260)\to J/\psi \pi^+ \pi^-) < 1.0$ at 90\% confidence level.
\end{abstract}

\maketitle

The surprising discovery of new states decaying to  $J/\psi \pi^+ \pi^-$
~\cite{charmonium,babar_Y} has renewed interest in the field of charmonium spectroscopy, as the new
states are not easy to accommodate in the quark
model.
In particular, the \babar\ experiment has discovered a new broad state, $Y(4260)$,
decaying to $J/\psi \pi^+ \pi^-$ in the initial-state-radiation (ISR)
reaction $e^+ e^- \to \gamma_{ISR} Y(4260)$. 
The quantum numbers  $J^{PC}=1^{--}$
are inferred from the single virtual-photon production mechanism.
Structure,
possibly related to the $Y(4260)$, has been observed in the reaction
$e^+ e^- \to \gamma_{ISR} \psi(2S) \pi^+ \pi^-$~\cite{babar_Y2}. 
A charmonium
state at this mass would be expected to decay predominantly to \DDb, \DDstarb or
\DstarDstarb~\cite{eichten}. 
It is peculiar that the decay rate to the hidden charm final state
$J/\psi \pi^+ \pi^-$ is much larger for the $Y(4260)$ than for excited
charmonium states~\cite{mo},
and that at the $Y(4260)$ mass the cross section for $e^+e^-\to\text{hadrons}$ 
exhibits a local minimum~\cite{pdg}.
Many theoretical interpretations for the $Y(4260)$ have been proposed, 
including unconventional scenarios:  
quark-antiquark gluon hybrids~\cite{Y4260-hybrid}, tetraquarks~\cite{maiani} 
and hadronic molecules~\cite{Y4260-molecule}. 
For a discussion and a list of 
references see, for example, Ref.~\cite{swanson}.

This work explores ISR production of the \DDb final state for evidence of charmonium states and 
unconventional structures. A study by the BELLE collaboration of 
the \DDstarb, and
\DstarDstarb final states can be found in Ref.~\cite{dstar2}.

This analysis is based on
 a 384~$\invfb$ data sample recorded at the
\FourS resonance and 40 MeV$/c^2$ below the resonance by the \babar\ detector at the \pep2
asymmetric-energy $e^+e^-$ storage rings.  
The \babar\ detector is
described in detail elsewhere~\cite{babar}.
Charged particles are detected
and their momenta measured by a combination of a 
cylindrical drift chamber (DCH)
and a silicon vertex tracker (SVT), both operating within a
1.5-T magnetic field of a superconducting solenoid. 
A ring-imaging Cherenkov detector (DIRC) combined with energy-loss measurements in the 
SVT and DCH are used to identify charged kaon and pion candidates. Photon energies are measured with a 
CsI(Tl) electromagnetic calorimeter (EMC).

$\DDb$ candidates are reconstructed in seven combinations of $D$ decay modes,
listed in Table~\ref{tab:list}~\cite{conj}. 
\begin{table*}[tbp]
\caption{List of the reconstructed final states and corresponding values of
  efficiency times branching fraction.}
\label{tab:list}
\begin{center}
\vskip -0.2cm
\begin{tabular}{lllc}
\hline \noalign{\vskip1pt}
Channel \ \ \ \  & First $D$ decay mode \ \ \   &   Second $D$ decay mode \ \ &
  $\epsilon_i^{\cal{B}}(m_{\DDb})$ ($\times 10^{-3}$) \cr
\hline \noalign{\vskip2pt}
1. $\Dz \Dzb$ & $\Dz \to \Km \pip$ & $\Dzb \to \Kp \pim$ & 0.14 \cr
2. $\Dz \Dzb$ & $\Dz \to \Km \pip$ & $\Dzb \to \Kp \pim \piz$ & 0.42 \cr
3. $\Dz \Dzb$ & $\Dz \to \Km \pip$ & $\Dzb \to \Kp \pim \pip \pim$ & 0.18
\cr
4. $\Dz \Dzb$ & $\Dz \to \Km \pip \piz$ & $\Dzb \to \Kp \pim \pip \pim$ &
0.26 \cr
5. $\Dp \Dm$ & $\Dp \to \Km \pip \pip$ &  $\Dm \to \Kp \pim \pim$ & 0.37
\cr
6. $\Dp \Dm$ & $\Dp \to \Km \pip \pip$ &  $\Dm \to  \Kp \Km \pim$ & 0.057
\cr
7. $\Dp \Dm$ & $\Dp \to \Km \pip \pip$ & $\Dm \to \KS \pim$ & 0.042 \cr
\hline \noalign{\vskip1pt}
\end{tabular}
\end{center}
\end{table*}
In each channel we allow any number of photons in the event.
Events are selected if the number of well-measured tracks is exactly
equal to the total number of charged daughter particles for the \D and the \Db
final states. 
Neutral pion candidates are formed 
from pairs of photons each having an energy greater than 30 MeV. The \KS
candidates are
reconstructed in the \pip\pim decay mode.
The tracks of each \D candidate are geometrically constrained to come from
a common vertex. Additionally, for the $\Dz \to \Km \pip \piz$ channel, 
\piz and \Dz mass constraints are 
included in the fit, and for the $\Dm \to \KS \pim$ channel a \KS mass
constraint is imposed. \D candidates with a $\chi^2$ fit probability greater than 0.1\% are retained.
Subsequently, each $\DDb$ pair is refitted to a common vertex with
the constraint that they originate from the $e^+ e^-$ interaction region;
only candidates with a $\chi^2$ fit probability greater than 0.1\% are retained.
Extra \piz candidates may originate from random combinations of photons.
Aside from \piz's from \Dz decays, we require that there
be no more than one other \piz candidate in the event
(except for channel 4, where we require that there are none). 

For \D channels without a \piz candidate, the \D  momentum is determined from the summed 
3-momenta of the decay particles and the energy is computed using the nominal \D mass value~\cite{pdg,cleo}.
For the $\Dz \to \Km \pip \piz$ channel, the 4-momentum from the mass
constrained fit is used.
This procedure gives similar $\DDb$ mass resolutions for all the channels.

The ISR photon, preferentially emitted at small angles 
with respect to the beam axis, escapes detection in approximately 90\% of
events. We therefore 
reconstruct the ISR photon as a missing particle. 
We define the squared recoil mass ($\MM$) to the $\DDb$ system using the four-momenta of the beam
particles ($p_{e^\pm}$) and the reconstructed $\D$ ($p_D$) and $\Db$ ($p_{\Db}$):
\begin{equation}
\MM \equiv ( p_{e^-}+p_{e^+}-p_D-p_{\Db})^2.
\end{equation}
This quantity 
should peak near zero for ISR events and for exclusive production of 
$e^+ e^- \to \DDb$ or $e^+ e^- \to \DDstarb$. In the latter
case, the \DDb mass distribution peaks at masses well above 6 GeV$/c^2$. Therefore we 
select ISR events by requiring a \DDb invariant mass below 6 GeV$/c^2$
and $\mid\MM\mid<1$ GeV$^2/c^4$.

Monte Carlo simulations of $e^+ e^- \to \gamma_{ISR}\DDb$ and candidates from the process
$e^+ e^- \to \gamma_{ISR}J/\psi$, $J/\psi \to K^+ K^- \pi^+ \pi^-$
in data are used to validate the requirement on the number of residual \piz and the shape 
of the $\MM$ distribution.

\begin{figure}[!htb]
\begin{center}
\includegraphics[width=8cm]{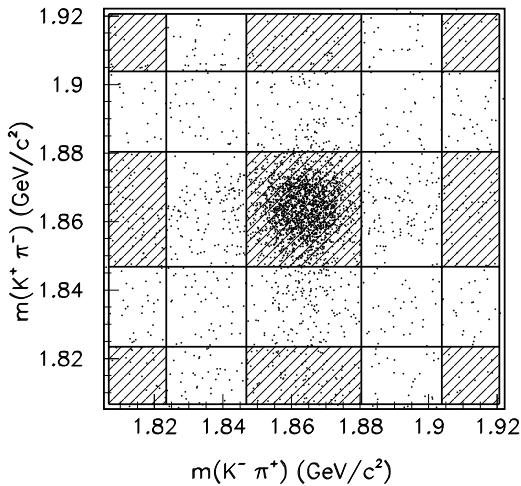}
\caption{$\Kp \pim$ mass vs. $\Km \pip$ mass distribution for final state 1. The
  cross-hatched areas correspond to the
  signal and sideband regions.}
\label{fig:fig0_dd}
\end{center}
\end{figure}

To estimate the number of background events
in the signal region, the two-dimensional space spanned by the invariant
masses of the two \D candidates in each event is divided into nine
regions: a central signal region and eight sideband regions above and 
below the signal regions, as illustrated in Fig.~\ref{fig:fig0_dd} for $\DDb$ candidates
reconstructed for the case of the $\Km \pip$ and $\Kp \pim$ modes. 
The mass range
for the signal region is within $\pm2.5\sigma$ of the $\D$ mass, and the 
sideband regions are $2.5\sigma$ wide and are separated from the signal
region by $3.5\sigma$, where $\sigma$ is the mass resolution determined from
a fit of a single Gaussian to the $\D$ candidate mass spectrum.

\begin{figure}[!htb]
\begin{center}
\includegraphics[width=8cm]{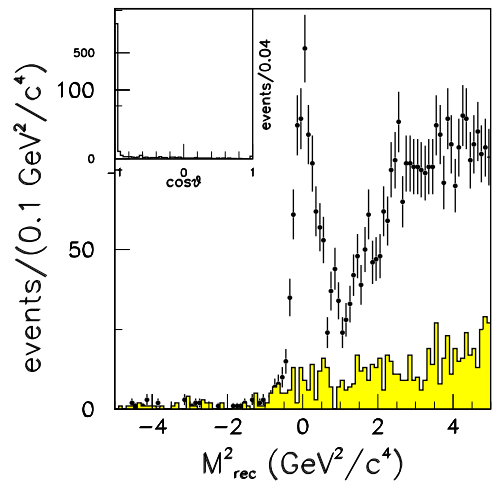}
\caption{Squared recoil mass, summed over all \DDb channels for ISR event candidates. 
The shaded histogram corresponds to non-\DDb 
background estimated from the \DDb-mass sidebands. The small inset shows the
distribution of the center-of-mass polar angle of the \DDb system in the ISR region.}
\label{fig:fig1_dd}
\end{center}
\end{figure}

The distribution of $\MM$, summed over all $\DDb$ channels,
is shown in Fig.~\ref{fig:fig1_dd}. The shaded histogram corresponds to the 
background in the signal region estimated from the \DDb mass sidebands.  
The
small inset in Fig.~\ref{fig:fig1_dd} shows the distribution of the \DDb 
center-of-mass polar
angle $\theta$ for \DDb candidates with $\mid\MM\mid<1$ GeV$^2/c^4$. 
The sharp peak at $cos \theta=-1$ is typical of ISR production
and agrees with Monte Carlo simulations.

\begin{figure}[!htb]
\begin{center}
\includegraphics[width=9.0cm]{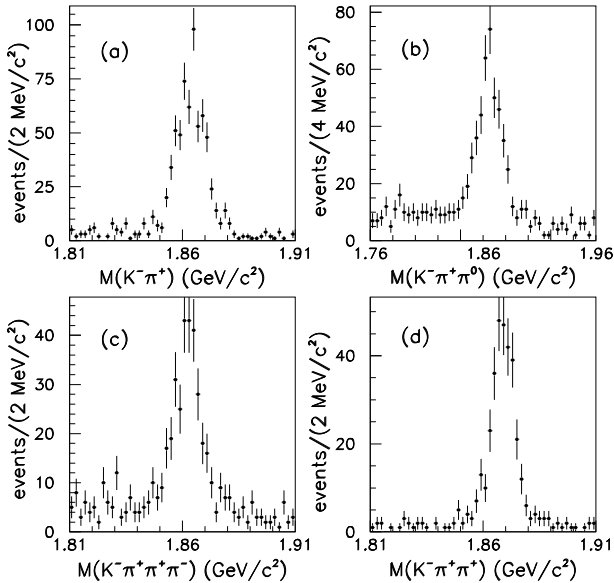}
\caption{$D$-candidate mass projections for events 
with \\ $\mid\MM\mid<1$ GeV$^2/c^4$ and 
a \DDb invariant mass below 6 GeV$/c^2$.
(a) $K^- \pi^+$ mass spectrum summed over channels 1, 2, and 3. (b) $K^- \pi^+
\pi^0$ mass spectrum summed over channels 2 and 4. (c) $K^- \pi^+ \pi^+ \pi^-$
mass spectrum summed over channels 3 and 4. (d)  $K^- \pi^+ \pi^+$ mass
spectrum
for channel 5.}
\label{fig:fig2_dd}
\end{center}
\end{figure}

The purity of each reconstructed $\D$ channel is demonstrated in 
Fig.~\ref{fig:fig2_dd} where
projections of the candidate $\D$ mass distribution for events
with $\mid \MM \mid < 1$ GeV$^2/c^4$
are shown.
Background is low in all channels.

\begin{figure}
\begin{center}
\vskip -0.15in
\includegraphics[width=10cm]{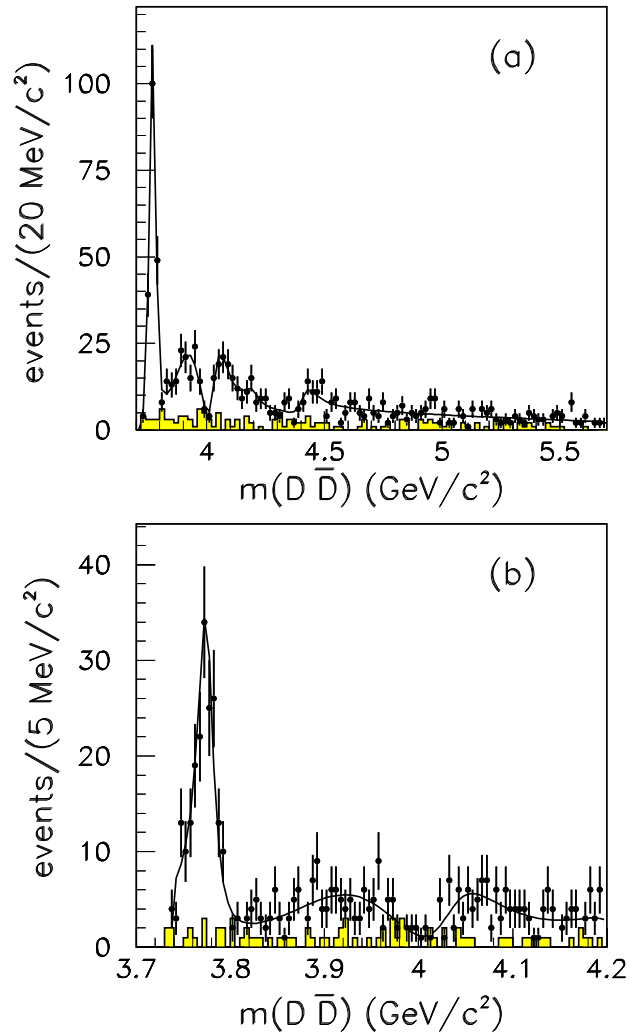}
\vskip -0.15in
\caption{(a) The ISR \DDb mass spectrum.
The shaded mass spectrum is from \DDb mass sidebands. The curve results
from the fit described in the text. (b) An expanded view of the region with $m_{\DDb}<4.2$ GeV$/c^2$.}
\label{fig:fig3_dd}
\end{center}
\end{figure}

The \DDb mass spectrum summed over all channels (860 events) is shown in
Fig.~\ref{fig:fig3_dd} where the curves are the results from the fit described
later. The
shaded histogram represents the background determined using the \DDb sideband
regions and corresponds to 17.5\% and 7.1\% of the signal candidates
for \DzDzb and \DpDm, respectively. We observe a clear \psiprpr signal and
other structures at the positions of $\psi(4040)$ and  $\psi(4415)$.  
We also observe a significant structure in the 3.9 \gevcc region, which may not
be due to a resonance; 
the coupled-channel model of Ref.~\cite{eichten1} in fact describes qualitatively the
observed \DDb mass spectrum and the structure around 3.9 GeV$/c^2$ without any
need for additional $\psi$ states.

To understand the background, we compute the expected contribution from ISR
production of the $D^{*0} \bar D^0$ system. Using Monte Carlo simulations and
the cross section estimate
from Ref.~\cite{dstar2} we find $\approx$ 6 \% as possible contamination.
This is confirmed by the examination of 
$D \gamma$ and $D \piz$ mass distributions where we find little evidence for
$D^{*0}$ signal. 
In contrast, strong evidence for $D^{*0}$ production is observed for $\MM >1.5$ GeV$^2/c^4$.
We investigate the possibility of background contributions from \DDbX final
states (where $X \ne \gamma$) by exploring events in the $\MM$ sideband region
$1.5 < \MM < 2.5$ GeV$^2/c^4$. The \DDb mass spectrum for these events
shows no structure. We conclude that the residual background to our signal is consistent with
originating mostly from combinatorial non-\DDb events.

In order to measure efficiency and \DDb mass resolution,
ISR events are simulated at eight different values of the $\DDb$ invariant
mass between 
3.75 and 7.25 GeV$/c^2$.
These events are generated using the
GEANT4 detector simulation package~\cite{geant} and are processed through the same
reconstruction and analysis chain as are real events. The mass-dependent efficiency 
for each channel is fitted using a second-order polynomial.
The mass resolution is determined from the difference between generated and
reconstructed $\DDb$ mass. 
The $\DDb$ mass resolution
is similar for all channels and increases with $\DDb$ mass from 1.5 to 5
MeV$/c^2$.
We observe good agreement between Monte Carlo and data $\MM$ distributions.

We define $N_i(m_{\DDb})$ as the number of \DDb candidates for channel $i$. The
channel branching fraction is $\BR_i$, 
and $\epsilon_i(m_{\DDb})$ is the efficiency as parametrized by the fitted polynomial.
We define as $\epsilon_i^{\cal{B}}(m_{\DDb})$ the product efficiency times
branching fraction for each channel,
\begin{equation}
\epsilon_i^{\cal{B}}(m_{\DDb}) = \epsilon_i(m_{\DDb}) \times \BR_i,
\end{equation}
and then compute $\epsilon^{\cal{B}}(m_{\DDb})$ as
\begin{equation}
\epsilon^{\cal{B}}(m_{\DDb}) = \frac{\sum_{i=1}^{7} N_i(m_{\DDb})}{\sum_{i=1}^{7} \frac{N_i(m_{\DDb})}{\epsilon^{\cal{B}}_i(m_{\DDb})}}.
\end{equation}
The values of $\epsilon_i^{\cal{B}}(m_{\DDb})$ are proportional to the expected yield for
each channel. Their values, integrated over the \DDb
mass spectrum, are reported in Table~\ref{tab:list}.
The resulting yields, corrected for efficiency and branching fractions,
are found to be consistent within the errors.

\begin{figure}[!htb]
\begin{center}
\includegraphics[width=9cm]{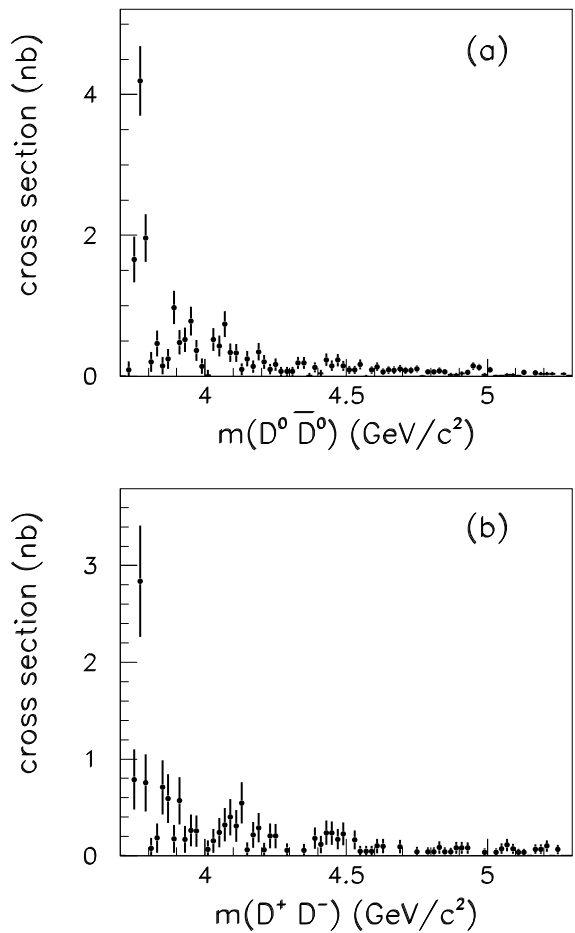}
\caption{(a) \DzDzb and (b) \DpDm cross sections with statistical uncertainties only.}
\label{fig:fig4_dd}
\end{center}
\end{figure}

The $\DDb$ cross section is computed using
\begin{equation}
\sigma_{e^+e^-\to \DDb}(m_{\DDb}) = \frac{dN/dm_{\DDb}}{\epsilon^{\cal{B}}(m_{\DDb})
  dL/dm_{\DDb}}. 
\end{equation}
The differential luminosity is computed as~\cite{benayoun}
\begin{equation}
\frac{dL}{dm_{\DDb}} = L \frac{2m_{\DDb}}{s} \frac{\alpha}{\pi
  x}(\ln(s/m_e^2)-1)(2-2x+x^2), 
\end{equation}
where $\alpha$ is the fine-structure constant, $x = 1 - m_{\DDb}^2/s$,
$s$ is the square of the 
$e^+ e^-$ center-of-mass energy,
$m_e$ is the electron mass, and $L$ is the
integrated luminosity of 384~\invfb.
The background-subtracted cross sections for \DzDzb and \DpDm, averaged over
20 \mevcc bins, are shown
in Fig.~\ref{fig:fig4_dd}.

Systematic errors on the cross sections (10.9\% for \DzDzb and  8.1\% for \DpDm)
include uncertainties in 
the particle identification efficiencies and tracking efficiency, 
possible inaccuracies in the
simulation of extraneous \piz candidates, and uncertainties in the background
estimates ($\approx$ 6 \%) and on the luminosity function ($\approx$ 1 \%). 
 
Integrating the cross sections in the \psiprpr region (3.74-3.80 \gevcc), 
we compute the ratio of branching fractions,
\begin{equation}
 \frac{\BR(\psiprpr \to  \DzDzb)}{\BR(\psiprpr \to \DpDm)} = 1.78 \pm 0.33
\pm 0.24,
\end{equation}
\noindent
to be compared with the value of $1.28 \pm 0.14$ reported by the PDG~\cite{pdg}.

We perform an unbinned maximum likelihood fit to the \DDb mass spectrum summed
over all channels. 
The parameters of the $\psi(4040)$,
$\psi(4160)$, and $\psi(4415)$ are fixed to the values reported in
Ref.~\cite{bes2} while the $Y(4260)$ parameters are taken from our measurement from the 
$J/\psi\pi^+ \pi^-$ channel \cite{babar_Y}.
The parameters of the \psiprpr are left free in the fit.
In addition, we search for evidence of the $Y(4260)$ in this spectrum.
Resolutions effects have been ignored since the widths of the resonances are
much larger than the experimental resolution. 

We express the total \DDb production as
\begin{equation}
f \left | P + c_1 W_1 e^{i \phi_1} + c_2 \sqrt{G} e^{i \phi_2} + ... + c_n W_n
  e^{i \phi_n} \right |^2 + (1 - f)B,
\end{equation}

\noindent
where $c_i$ and
$\phi_i$ are free parameters, $W_i$ are spin-1 relativistic Breit-Wigner
distributions, $P$ represents the non-resonant contribution, $B$ describes
the non-\DDb background and $f$ (0.829 $\pm$ 0.015) is the signal fraction.
The efficiency $\epsilon^{\cal{B}}(m_{\DDb})$ is almost linear and 
increases from $\approx 2 \times 10^{-3}$ to $\approx 4 \times 10^{-3}$ in 
the fitted mass region. It has been 
parametrized by a $2^{nd}$ order polynomial and it has been multiplied by 
$P$ and $W_i$.
The data require that we include the 3.9 GeV$/c^2$ structure,
as suggested in Ref.~\cite{eichten1},
which we parameterize empirically as the square root of
a Gaussian times a phase factor ($\sqrt{G} e^{i \phi_2}$). 
The parameters of the Gaussian are left free, and the phase 
allows interference with the $\psi$ states.

We find that, in order to have a satisfactory description 
of the data, interference must be allowed between the resonances and the
non-resonant contribution $P$. The latter contribution is parametrized 
either as a linear ($a + b m$) or a threshold function 
$(m-m_{th})^a e^{-bm -cm^2}$, where $m=m_{\DDb}$,
$m_{th}$ is the threshold \DDb mass, and $a$, $b$ and $c$ are free parameters.
This threshold function has also been used to describe the non-\DDb background 
$B$.

The two
different parametrizations give similar results, which are considered in the evaluation
of the systematic errors. These include also uncertainties in the $D$ mass 
and on the overall $\DDb$ mass scale.
The size of the non-resonant production is determined by the fit.
 
The fit with a linear non-resonant contribution is shown in
Fig.~\ref{fig:fig3_dd}(a). 
Figure ~\ref{fig:fig3_dd}(b) shows
an expanded view of the threshold region. 

The fit returns the following parameters for the $G(3900)$ structure and for the \psiprpr:
\begin{equation}
 m(G(3900)) = (3943  \pm 17_{stat} \pm 12_{syst}) \mevcc,
\end{equation}
\begin{equation}
\sigma(G(3900))=(52 \pm 8_{stat} \pm 7_{syst}) \mevcc, 
\end{equation}
\begin{equation}
 m(\psiprpr) = (3778.8 \pm 1.9_{stat} \pm 0.9_{syst}) \mevcc, 
\end{equation}
\begin{equation}
 \Gamma(\psiprpr) = (23.5 \pm 3.7_{stat} \pm 0.9_{syst} ) \mev.
\end{equation}

The systematic error on the \psiprpr mass includes uncertainties in the
$D$ mass, background
parametrization, and detector related issues such as magnetic field, EMC
corrections and energy loss.
We measure a significantly higher \psiprpr mass with respect to previous
measurements ($3772.4 \pm 1.1$) $\mevcc$~\cite{pdg}.
The change in likelihood due to the inclusion of a $Y(4260)$ amplitude in the
fit is 
given by $\Delta(2 \ln(L)) = 0.1$ with 
two additional fit parameters. 

The systematic errors due to the masses and the widths of the $\psi(4040)$,
$\psi(4160)$, $\psi(4415)$ and $Y(4260)$
resonances in the fit
are evaluated by varying them by their statistical uncertainties. 
The signal fraction has been varied within its statistical error and the
meson radius used in the Blatt-Weisskopf damping factor~\cite{dump} present 
in the relativistic Breit-Wigner has been varied between 0 and 5 ${\rm GeV}^{-1}$.
The deviations from the central values are 
added in quadrature. The uncertainty on $\epsilon^{\cal{B}}(m_{\DDb})$ is
evaluated by using a weighted mean of branching 
fraction and efficiency uncertainties for the different channels.
The fitted $Y(4260)$ yield before efficiency correction is 
$0.2 \pm 6.1_{stat} \pm 2.8_{syst}$ events.

This $Y(4260)$ yield in the $\DDb$ channel is used to 
compute the cross
section times branching fraction, which can then be compared to our measurement from the 
$J/\psi\pi^+ \pi^-$ channel \cite{babar_Y}.
We obtain
\begin{equation}
\frac{\BR(Y(4260)\to \DDb)}{\BR(Y(4260)\to J/\psi \pi^+ \pi^-)} < 1.0,
\end{equation}
or
\begin{equation}
\Gamma(Y(4260) \to e^+ e^-) \cdot \BR(Y(4260)\to \DDb) < 5.7 \ev,
\end{equation}
\noindent
at 90\% confidence level. 

In conclusion, we have studied the exclusive ISR production of the \DDb system.
The mass spectrum is dominated by $J^{PC}=1^{--}$ states; in particular the \psiprpr
is clearly seen.
In order to fit the mass spectrum, signals from $\psi(4040)$, $\psi(4160)$, and $\psi(4415)$ have been
included. The fit requires the presence of a broad structure near 3900 \mevcc. The presence of an enhancement in this
region is predicted by a coupled channel model from Eichten {\it et
  al.}~\cite{eichten1},  although the possibility of the presence of a new
$\psi$ state cannot be excluded.

If the $Y(4260)$ is a $1^{--}$ charmonium state, it should decay predominantly to 
$\DDb$~\cite{eichten}; however no evidence is found for $Y(4260)$ decays to $\DDb$.
Other explanations have been proposed, such as a
hybrid, baryonium, or tetraquark state. 

We are grateful for the excellent luminosity and machine conditions
provided by our \pep2\ colleagues,
and for the substantial dedicated effort from
the computing organizations that support \babar.
The collaborating institutions wish to thank
SLAC for its support and kind hospitality.
This work is supported by
DOE
and NSF (USA),
NSERC (Canada),
CEA and
CNRS-IN2P3
(France),
BMBF and DFG
(Germany),
INFN (Italy),
FOM (The Netherlands),
NFR (Norway),
MIST (Russia),
MEC (Spain), and
STFC (United Kingdom).
Individuals have received support from the
Marie Curie EIF (European Union) and
the A.~P.~Sloan Foundation.

\renewcommand{\baselinestretch}{1}


\begin{thebibliography}{99}
\bibitem{conj}
Charge conjugate states are implied throughout this work.
\end{thebibliography}
\end{document}